\documentclass[twocolumn,showpacs,floatfix,prl]{revtex4}%
\usepackage{graphicx}%
\usepackage{amsmath}%
\usepackage{amsfonts}%
\usepackage{amssymb}
\usepackage{bm}

\def\k{{ {\bm k} }}

\def\q{{ {\bm q} }}
\def\Q{{ {\bm Q} }}

\def\0{{ {\bm 0} }}
\def\w{{\omega}}
\def\a{{\alpha}}

\def\g{{\gamma}}

\allowdisplaybreaks[4]

\begin{document}
\title{
High-$T_{\rm c}$ Superconductivity near the
Anion Height Instability in \\ 
Fe-based Superconductors: Analysis of LaFeAsO$_{1-x}$H$_x$
}
\author{
Seiichiro \textsc{Onari}$^{1}$, 
Youichi \textsc{Yamakawa}$^{2}$, and
Hiroshi \textsc{Kontani}$^{2}$
}

\date{\today }

\begin{abstract}
The isostructural transition in the tetragonal phase,
with sizable change in the anion-height, is realized 
in heavily H-doped LaFeAsO and (La,P) co-doped CaFe$_2$As$_2$.
In these compounds, the superconductivity with higher-$T_{\rm c}$ ($40\sim50$K)
is realized near the isostructural transition.
To find the origin of the anion-height instability 
and the role in realizing the higher-$T_{\rm c}$ state,
we develop the orbital-spin fluctuation theory
by including the vertex correction.
We analyze LaFeAsO$_{1-x}$H$_x$, and find that the non-nematic 
orbital fluctuations, which induce the anion-height instability, 
are automatically obtained at $x\sim0.5$,
in addition to the conventional nematic orbital fluctuations at $x\sim0$.
The non-nematic orbital order triggers the isostructural transition,
and its fluctuation would be a key ingredient to
realize higher-$T_{\rm c}$ superconductivity of order $50$K.

\end{abstract}

\address{
$^1$ Department of Applied Physics, Nagoya University,
Furo-cho, Nagoya 464-8603, Japan. 
\\
$^2$ Department of Physics, Nagoya University,
Furo-cho, Nagoya 464-8602, Japan. 
}
 

\sloppy

\maketitle


The normal-state phase diagram of Fe-based superconductors
is important to reveal the essential electronic states
and the mechanism of superconductivity.
In many compounds, the structure transition from tetragonal ($C_4$)
to orthorhombic ($C_2$) is realized at $T_{\rm S}$, and the 
antiferromagnetic (AFM) order appears at $T_{\rm N}$ below $T_{\rm S}$.
The superconductivity is realized near the structural 
quantum critical point (QCP) at $T_S=0$ and/or the magnetic QCP at $T_N=0$.
For example, the optimum $T_{\rm c}$ in FeSe$_x$Te$_{1-x}$
is realized near the structural QCP at $x\approx0.6$ \cite{FeSe1},
whereas AFM order is absent for $x>0.5$.

To explain the $C_2$ structure transition,
both the spin-nematic \cite{Fernandes} and 
orbital-nematic \cite{Kruger,PP,WKu,Onari-SCVC,Dagotto}  
mechanisms had been proposed.
In the latter scenario, 
orbital-nematic order is induced by spin fluctuations,
due to strong orbital-spin mode-coupling described by the 
vertex correction (VC) \cite{Onari-SCVC}.
Both mechanisms can explain the shear modulus $C_{66}$ softening 
\cite{Yoshizawa,Bohmer}.
The orbital mechanism would be consistent with the 
large $d$-level splitting $E_{yz}-E_{xz}\sim500$K 
in the $C_2$ phase \cite{ARPES-Shen,Shimojima},
and with the large orbital susceptibility given by Raman spectroscopy
\cite{Gallais,Kontani-Raman}.
The nematic order is observed by
the magnetic torque measurements \cite{Kasahara}.
Since the superconductivity is realized next to the orbital 
and spin ordered phases,
both fluctuations would be essential for the pairing mechanism.

However, this is not the whole story of Fe-pnictides:
The unique phase diagram of LaFeAsO$_{1-x}$H$_x$ 
with double-dome superconducting phase \cite{Iimura,Fujiwara}
attracts great attention.
The second superconducting dome ($x\ge0.2$) 
is next to the ``$C_4$ isostructural phase transition'' 
with sizable change in the $c$-axis length (or anion-height) 
for $0.45<x<0.5$
\cite{Hosono-private,comment}. 
(The $c$-axis length is unchanged in the 
$C_2$ structure transition at $x\sim0$.)
Similarly, high-$T_{\rm c}$ ($\sim50$K) superconductivity is realized
near the ``collapsed $C_4$ phase'' in rare-earth doped CaFe$_2$As$_2$ 
\cite{collapsed,Nohara}.
In (La,P) co-doped CaFe$_2$As$_2$,
higher-$T_{\rm c}$ state is realized near the anion-height instability,
whereas it avoids the AFM phase as clearly shown in Ref. \cite{collapsed,Nohara}.
These experiments strongly indicate that 
the anion-height instability is a key ingredient 
for higher-$T_{\rm c}$ superconductivity of order 50K.
Authors in Ref. \cite{Avci} discussed that the $C_4$ phase 
in (Ba,Na)Fe$_2$As$_2$ originates from the $C_4$ magnetic order.
However, stripe magnetic order (=$C_2$ symmetry) is realized in 
LaFeAsO$_{1-x}$H$_x$ at $x\sim0.5$ \cite{Hosono-private},
which indicates small spin-lattice coupling.

\begin{figure}[!htb]
\includegraphics[width=.75\linewidth]{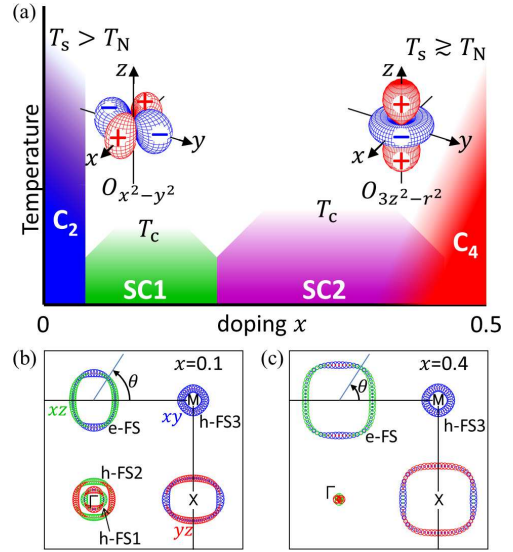}
\caption{(color online)
(a) Schematic phase diagram of LaFeAsO$_{1-x}$H$_x$.
We predict that non-nematic $O_{3z^2-r^2}$ (nematic $O_{x^2-y^2}$) 
charge quadrupole order emerges in the $C_4$ ($C_2$) phase.
(b)(c) FSs at $x=0.1$ and $x=0.4$ \cite{Yamakawa-H}.
e-FS is the electron-pocket, and h-FS3 (h-FS1,2) is the hole-pocket 
composed of $d_{xy}$ ($d_{xz},d_{yz}$) orbital.
}
\label{fig:QO-fig}
\end{figure}

In this paper,
we discuss the origin of the anion-height instability 
and its role of higher-$T_{\rm c}$ superconductivity.
For this purpose, we study LaFeAsO$_{1-x}$H$_x$ ($x=0\sim0.5$)
by calculating both the VC and the self-energy $\Sigma$ self-consistently.
By this ``self-consistent VC+$\Sigma$ (SC-VC$_\Sigma$) method'',
we obtain the non-nematic orbital order $O_{3z^2-r^2}$ at $x\sim0.5$.
This order parameter couples to the anion-height change
and triggers the $C_4$ isostructural transition,
which cannot be explained by the spin-fluctuation theories 
without the VC \cite{Yamakawa-H,Suzuki-H,Suzuki-H2}.
We also find that 
orbital-fluctuation-mediated $s$-wave state 
is stabilized by including the VC for the gap equation,
which is dropped in conventional Migdal-Eliashberg theory.
The present study reveals that
multiple kinds of orbital fluctuations play significant roles
in Fe-based superconductors.

Figure \ref{fig:QO-fig} (a) shows the phase diagram of LaFeAsO$_{1-x}$H$_x$:
We propose that the charge quadrupole order 
$O_{3z^2-r^2}\equiv \frac12 (n_{xz}+n_{yz})-n_{xy}$
($O_{x^2-y^2}\equiv n_{xz}-n_{yz}$) appears at $x\sim0.5$ ($x\sim0$).
The softening of the longitudinal modulus along the $c$-axis,
$C_{33}$, observed in under- and over-doped 
Ba(Fe$_{1-x}$Co$_x$)$_2$As$_2$ \cite{Yoshizawa-C33}
indicates that $O_{3z^2-r^2}$ quadrupole fluctuations exist
in various Fe-based compounds.

The tight-binding model of LaFeAsO$_{1-x}$H$_x$ for $0\le x\le0.5$
had been introduced by the present authors in Ref. \cite{Iimura}.
The Fermi surfaces (FSs) for $x=0.1$ and 0.4 are shown in 
Fig. \ref{fig:QO-fig} (b) and (c), respectively.
The intra-orbital nesting and inter-orbital one are 
the driving forces of the magnetic and orbital fluctuations, respectively.
We analyze the multiorbital Hubbard model 
with intra (inter) orbital interaction $U$ ($U'$) and 
the exchange interaction $J$ under the constraint $U=U'+2J$,
assuming uniform states.
Electronic phase separation due to the 
imperfect nesting is discussed in Ref. \cite{PhaseSeparation}.

Here, we denote $d_{3z^2-r^2}$, $d_{xz}$, $d_{yz}$, $d_{xy}$, $d_{x^2-y^2}$ 
orbitals as $1,2,3,4,5$.
The FSs are mainly composed of 2,3,4 orbitals.
The charge (spin) susceptibility 
${\hat \chi}^{c(s)}(q)$ is given in the $5^2\times5^2$ matrix form 
in the orbital basis as follows:
\begin{eqnarray}
{\hat \chi}^{c(s)}(q)= 
{\hat \Phi}^{c(s)}(q)(1-{\hat \Gamma}^{c(s)}{\hat \Phi}^{c(s)}(q))^{-1}
\label{eqn:chisc}
\end{eqnarray}
where $q=(\q,\w_l)$ 
and ${\hat \Phi}^{c(s)}(q)={\hat \chi}^{(0)}(q)+{\hat X}^{c(s)}(q)$:
${\hat \chi}^{(0)}(q)$ is the bubble susceptibility with self-energy correction,
and ${\hat X}^{c(s)}(q)$ is the VC for charge (spin) channel.
${\hat \Gamma}^{c(s)}$ is the matrix form of the bare Coulomb interaction 
for the charge (spin) sector \cite{Kontani-RPA}. 
In the original SC-VC$_\Sigma$ method, the VC is given by
the Maki-Thompson (MT) and Aslamazov-Larkin (AL) terms,
which are the first and second order terms with respect to 
${\hat \chi}^{c,s}$, respectively.
Since ${\hat X}^{c}\gg{\hat X}^{s}$ near the QCP,
we put ${\hat X}^{s}(q)=0$, and calculate only the
AL term for ${\hat X}^{c}(q)$ self-consistently.
Its justification is verified in Refs. \cite{Onari-SCVC,Ohno-SCVC},
and also confirmed by the 
recent renormalization group study \cite{Tsuchiizu}.

The charge (spin) Stoner factor $\a_{c(s)}$ is given by the maximum
eigenvalue of ${\hat \Gamma}^{c(s)}{\hat \Phi}^{c(s)}(q)$
in eq. (\ref{eqn:chisc}), and 
$\a_{c(s)}=1$ corresponds to the orbital (spin) order.
Although the relation $\a_{s}\gg\a_c$ is satisfied within the RPA for $J>0$,
the opposite relation can be realized 
if the VC is taken into account beyond the RPA.
Here, we introduce the quadrupole susceptibilities:
\begin{eqnarray}
\chi_\gamma^Q(\q,\w_l)= \sum_{l,l',m,m'}
O_\gamma^{l,l'}\chi^c_{l,l';m,m'}(\q,\w_l)O_\gamma^{m',m},
\end{eqnarray}
where $\gamma=x^2-y^2$, $3z^2-r^2$, $xz$, $yz$, $xy$ 
represents the quadrupole \cite{comment2}.
Then, $\chi^Q_{x^2-y^2}(q)\approx 
\chi^c_{2,2;2,2}(q)+\chi^c_{3,3;3,3}(q)-2\chi^c_{2,2;3,3}(q)$,
and 
$\chi^Q_{3z^2-r^2}(q)\approx \chi^c_{4,4;4,4}(q)
-\sum_{l=2,3}\chi^c_{l,l;4,4}(q)+\sum_{l,m=2,3}\chi^c_{l,l;m,m}(q)/4$.

Now, we study the tight-binding Hubbard models of LaFeAsO$_{1-x}$H$_x$
based on the SC-VC$_\Sigma$ method,
in which  both the VC and the one-loop self-energy $\hat{\Sigma}$ are
calculated self-consistently.
By this method, the mass-enhancement factor for $l$-orbital
is given as $1/z_l = 1-{\rm Re}d\Sigma_l(k,\w)/d\w|_{\w=0}$,
and we obtain 
$1/z_l=3\sim5$ for $l=2\sim4$ and $1/z_4>1/z_{2,3}$ in LaFeAsO$_{1-x}$H$_x$.
The expressions of the VC and $\hat{\Sigma}$ are explained in Refs. 
\cite{LiFeAs} in detail.
Hereafter, we fix the parameters $J/U=0.14$ and $T=0.05$eV,
and the unit of energy is eV.

\begin{figure}[!htb]
\includegraphics[width=.9\linewidth]{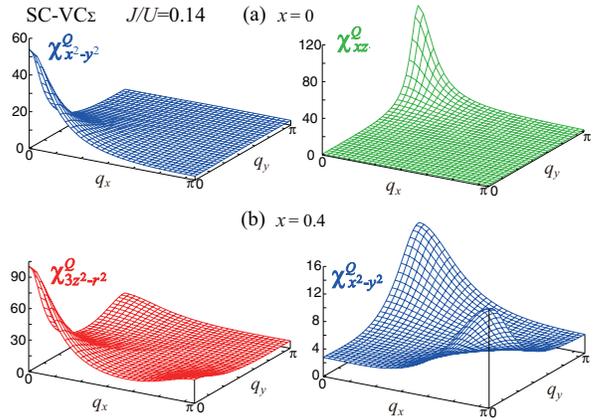}
\caption{(color online)
$\chi_\gamma^Q(\q)$ at zero frequency
obtained by the SC-VC$_\Sigma$ method:
(a) $\gamma=x^2-y^2$ and $\gamma=xz$ for $x=0$ ($U=2.06$), and
(b) $\gamma=3z^2-r^2$ and $\gamma=x^2-y^2$ for $x=0.4$ ($U=1.65$).
Note that $\chi_{xz}^Q(q_x,q_y)=\chi_{yz}^Q(q_y,q_x)$.
Similar results are obtained by the SC-VC method \cite{suppl}.
}
\label{fig:chiQ-U}
\end{figure}

Figure \ref{fig:chiQ-U} shows the largest two static
quadrupole susceptibilities $\chi^Q_\gamma(\q)$ 
for (a) $x=0$ and (b) $x=0.4$, respectively.
For each $x$, the relations $\a_c=0.97$ and $a_c> \a_s\sim0.9$ are satisfied,
consistently with the relation $T_S>T_N$.
At $x=0$ in (a), we obtain the strong developments of
$\chi_{x^2-y^2}^Q({\bm 0})$ and $\chi_{xz}^Q({\bm Q})$,
similarly to the previous SC-VC analysis \cite{Onari-SCVC}.
The divergence of $\chi_{x^2-y^2}^Q({\bm 0})$ 
causes the $C_2$ structure transition.
In addition, large antiferro-orbital fluctuations 
are induced by the cooperation of the VC and the good inter-orbital nesting
between e-FS and h-FSs \cite{Onari-SCVC}.
The shear modulus 
$C_{66}\propto 1-g_{x^2-y^2}\chi^Q_{x^2-y^2}(\bm{0})$ reaches zero
even if $\chi^Q_{x^2-y^2}(\bm{0})$ in the SC-VC$_\Sigma$ method is finite, 
where $g_{x^2-y^2}(\ll1)$ is the quadrupole interaction
due to the acoustic phonon \cite{Kontani-softening}.

At $x=0.4$ in Fig. \ref{fig:chiQ-U} (b), in contrast,
we obtain the large peak of $\chi_{3z^2-r^2}^Q({\bm 0})$ due to the VC.
Since its divergence induces the change in the ratio
$n_{xy}/n_{xz}$ while keeping $n_{xz}=n_{yz}$,
the obtained large $\chi_{3z^2-r^2}^Q({\bm 0})$
gives the non-nematic ($C_4$) orbital fluctuations 
and anion-height instability.
In addition, large antiferro-orbital fluctuations 
$\chi_{x^2-y^2}^Q({\bm Q})$ are also induced by the VC.
It is noteworthy that a strong interorbital charge transfer
from in-plane to out-of-plane orbitals is observed in Co-doped BaFe$_2$As$_2$
\cite{Ma}.

\begin{figure}[!htb]
\includegraphics[width=.7\linewidth]{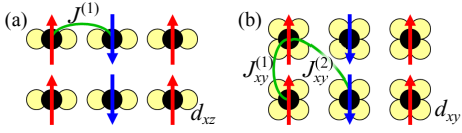}
\caption{(color online)
(a) Localized $(d_{xz},d_{yz})$-orbital model with KK coupling. 
(b) Localized $d_{xy}$-orbital model with Heisenberg coupling.
Here, the occupied orbitals are shown.
}
\label{fig:KK}
\end{figure}

Here, we try to understand the orbital-spin mode-coupling
due to AL term in terms of the localized picture $U\gg W_{\rm band}$:
First, we introduce the Kugel-Khomskii (KK) type orbital-dependent
exchange interaction \cite{Kugel}
between the nearest neighbor $d_{xz},d_{yz}$-orbitals,
$H'\sim J^{(1)}\sum_{\langle i,j \rangle}^{\rm N.N}
({\bm s}_i \cdot{\bm s}_j)(n_{xz}^i n_{xz}^j \delta_{i-j,(\pm1,0)}
+n_{yz}^i n_{yz}^j \delta_{i-j,(0,\pm1)})$,
as shown in Fig. \ref{fig:KK} (a).
Note that $J^{(1)}\sim 2t^2/U$.
Due to this orbital-spin coupling term, 
if the AFM order with ${\bm Q}=(\pi,0)$ is realized,
the electrons at each site will occupy the $d_{xz}$-orbital,
as shown in Fig. \ref{fig:KK} (a).
That is, the AFM order or fluctuations induces the 
$C_2$ orbital order ($n_{xz}\ne n_{yz}$) or fluctuations, 
and vice versa.
Next, we consider the single $d_{xy}$-orbital model
with the nearest- and next-nearest-neighbor exchange interactions:
$H'' \sim J^{(1)}_{xy} \sum_{\langle i,j \rangle}^{\rm N.N}
({\bm s}_{i}\cdot{\bm s}_{j})(n_{xy}^i n_{xy}^j)
+J^{(2)}_{xy} \sum_{\langle i,j \rangle}^{\rm N.N.N}
({\bm s}_{i}\cdot{\bm s}_{j})(n_{xy}^i n_{xy}^j)$.
When $J^{(2)}_{xy} > \frac12 J^{(1)}_{xy}$,
the  $\Q=(\pi,0)$ AFM state in Fig. \ref{fig:KK} (b)
appears due to ``order-by-disorder'' mechanism
\cite{order-by-disorder}.

Now, we consider the three-orbital model $H'+H''$:
When  $J^{(2)}_{xy}\gg J^{(1)}$,
the ferro-orbital polarization $n_{xy} \gg n_{xz}=n_{yz}$ with AFM order
shown in Fig. \ref{fig:KK} (b)
would be realized to gain the exchange energy.
In this case, the AFM order or fluctuations induces non-nematic 
$C_4$ orbital order or fluctuations, and vice versa.
This case corresponds to $x\sim0.5$ with strong 
$d_{xy}$-orbital spin fluctuations.
Thus, the KK-type spin-orbital coupling is understandable
in term of the weak-coupling approach by including the AL term.
The strong coupling approaches are useful to 
understand the ordered phases \cite{strongcoupling}.

We also discuss why the VC induces the 
$C_4$ ($C_2$) order at $x=0.5$ ($x=0$) analytically:
When spin fluctuations develop mainly in the $l$-orbital,
the charge AL-term $X^c_{l,l;l,l}({\bm 0})\sim T\sum_\k \{\chi^s_{l,l;l,l}(k)\}^2$
becomes large \cite{Onari-SCVC,Ohno-SCVC}.
Now, we analyze $\chi_\g^Q({\bm0})$ 
by inputting only three irreducible susceptibilities
$\Phi_l^c \equiv \chi_{l,l;l,l}^{(0)}(\0)+ X^c_{l,l;l,l}(\0)$ ($l=2\sim4$) 
into eq. (\ref{eqn:chisc}).
For $J=0$, for simplicity, we obtain \cite{Ohno-SCVC,suppl}
\begin{eqnarray}
&&\chi_{x^2-y^2}^Q({\bm0}) = 2\Phi_2^c(1-U\Phi_2^c)^{-1},
\label{eqn:ap-chiQx2y2} \\
&&\chi_{3z^2-r^2}^Q({\bm0}) = b(1-aU\Phi_4^c)^{-1},
\label{eqn:ap-chiQz2}
\end{eqnarray}
where $a\equiv(5U\Phi_2^c-1)/(3U\Phi_2^c+1)$ and
$b\sim (5U\Phi_4^c+1)^2/16U^2\Phi_4^c$ near the QCP.
In the case of $\Phi_2^c=\Phi_3^c> a\Phi_4^c$, 
then $\chi_{x^2-y^2}^Q({\bm0})$ is the most divergent.
In the opposite case,
$\chi_{3z^2-r^2}^Q({\bm0})$ is the most divergent if $a$ is positive.
At $x\sim0.4$, h-FS1 and h-FS2 almost disappear
as shown in Fig. \ref{fig:QO-fig} (c), so 
$d_{xy}$-orbital spin fluctuations becomes dominant \cite{Yamakawa-H}. 
For this reason, at $x\sim0.4$,
the $O_{3z^2-r^2}$ order and anion-height instability are driven by 
$\Phi_4^c$ due to strong $d_{xy}$-orbital spin fluctuations.

\begin{figure}[!htb]
\includegraphics[width=.9\linewidth]{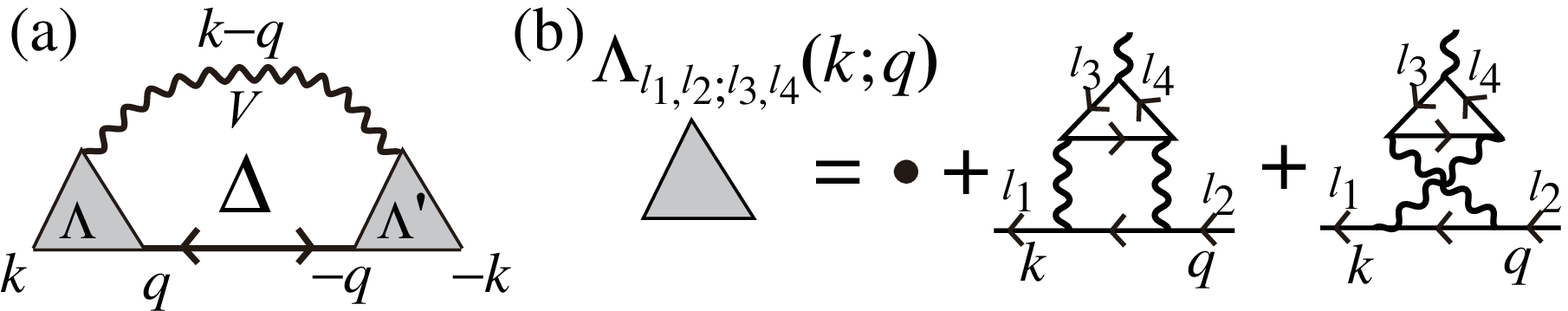}
\includegraphics[width=.9\linewidth]{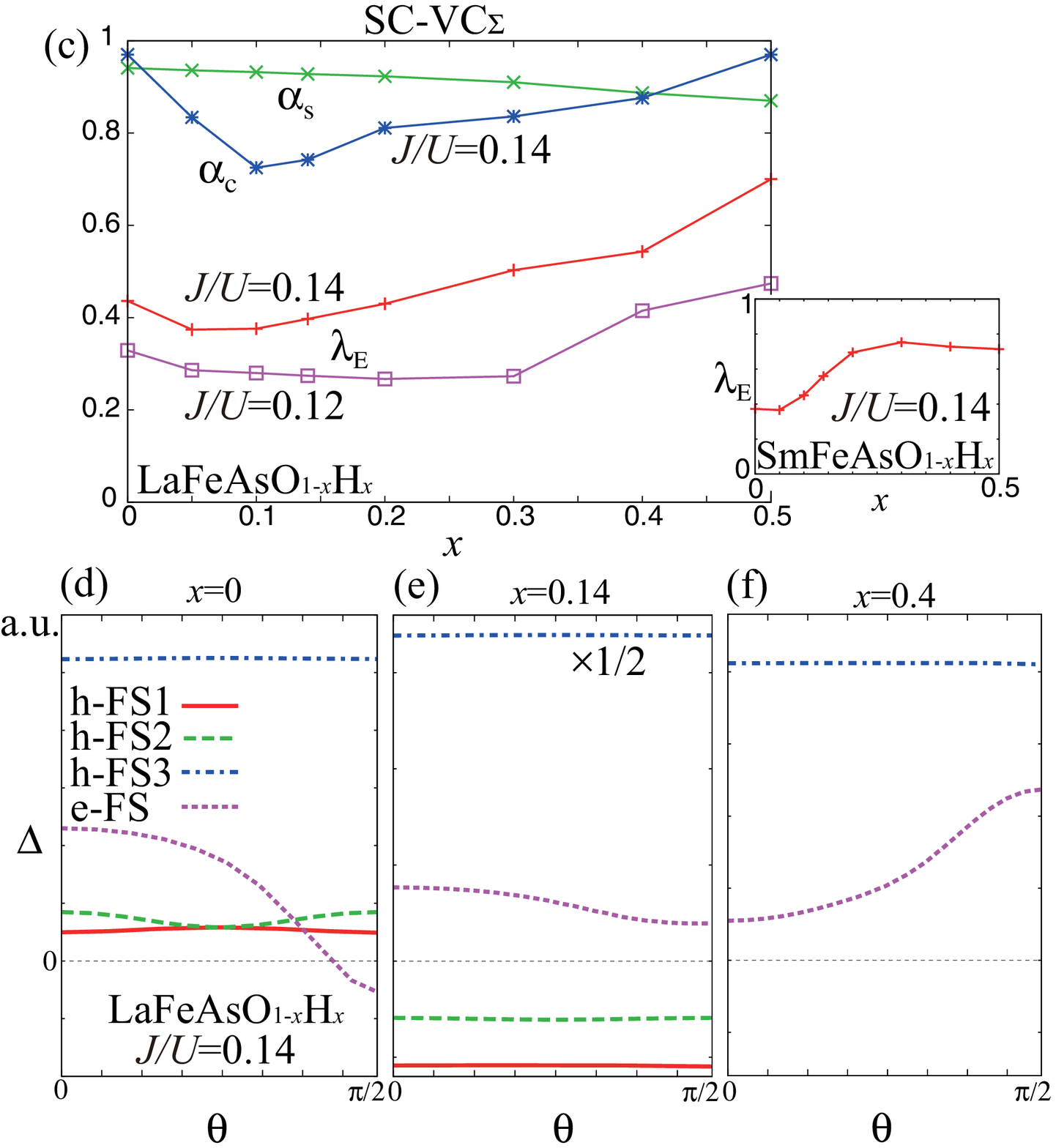}
\caption{(color online)
(a) Gap equation with $\Delta$-VC and (b) AL-type diagram for $\Lambda$.
(c) $\a_{c,s}$ and $\lambda_E$ as functions of $x$ in LaFeAsO$_{1-x}$H$_x$
for $J/U=0.14$.
$\lambda_E$ for $J/U=0.12$ is also shown.
(Inset: $\lambda_E$ in SmFeAsO$_{1-x}$H$_x$.)
The gap functions on the FSs  
at (d) $x=0$, (e) $x=0.14$ and (f) $x=0.4$ for $J/U=0.14$.
$\theta$ is the azimuthal angle in Fig. \ref{fig:QO-fig} (b) and (c).
}
\label{fig:SCgap}
\end{figure}

Now, we study the superconductivity
due to orbital and spin fluctuations,
based on the SC-VC$_\Sigma$ method.
In almost all previous studies, the VC for the 
gap equation ($\Delta$-VC) had been dropped.
In strongly correlated systems, however, 
$\Delta$-VC could be quantitatively important since the Migdal's theorem 
is not valid any more.
Since the AL-type VC for $\chi^Q_\gamma(q)$ is very large,
$\Delta$-VC due to AL-type diagram should be significant.
Here, we solve the following gap equation in the orbital-basis
by taking the $\Delta$-VC into account:
\begin{eqnarray}
\lambda_E\Delta_{l,l'}(k)&=& -T\sum_{q,m_i}V_{l,m_1;m_4,l'}(k,q) G_{m_1,m_2}(q)
\nonumber \\
& &\times \Delta_{m_2,m_3}(q)G_{m_4,m_3}(-q)
\end{eqnarray}
where $\lambda_E$ is the eigenvalue, 
$\Delta_{l,l'}(k)$ is the gap function, and
$G_{l,l'}(q)$ is the Green function with self-energy.
The pairing interaction $V_{l,m_1;m_4,l'}(k,q)$ is given as
\begin{eqnarray}
&&{\hat V}(k,q)= \frac32 {\hat \Lambda}^s(k,q)
{\hat \Gamma}^s{\hat \chi}^s(k-q){\hat \Gamma}^s
{\hat \Lambda}'^s(-k,-q)
\nonumber \\
&&\ \ \ \ \ 
-\frac12 {\hat \Lambda}^c(k,q)
{\hat \Gamma}^c{\hat \chi}^c(k-q){\hat \Gamma}^c
{\hat \Lambda}'^c(-k,-q) +V^{(1)}
 \label{eqn:V} 
\end{eqnarray}
where ${\hat \Lambda}^{c(s)}(k,q)$ is the vertex
for the charge (spin) channel shown in Fig. \ref{fig:SCgap} (a),
$\Lambda'^{c(s)}_{l,l';m,m'}(k,q)=\Lambda^{c(s)}_{m',m;l',l}(k,q)$,
and $V^{(1)}=\frac12({\hat \Gamma}^s-{\hat \Gamma}^c) \sim U$.
To make consistency with the SC-VC$_\Sigma$ method,
we calculate the AL-type contribution to ${\hat \Lambda}^{c}(k,q)$
given in Fig. \ref{fig:SCgap} (b),
whereas we put ${\hat \Lambda}^{s}(k,q)={\hat 1}$.

To study the superconducting state for $x=0\sim0.5$,
we introduce ${\bar U}(x)$ by the linear interpolation between 
$U_c=2.06$ at $x=0$ and $U_c=1.55$ at $x=0.5$, 
as done in Ref. \cite{Yamakawa-H}.
The obtained ${\bar U}(x)$ decreases with $x$, 
which will be given by the change in
the Kanamori screening, which is dropped in the 
present one-loop $\Sigma$.
In fact, the density of states at the Fermi level,
$N(0)$, increases by $30\%$, by changing $x$ from 0 to 0.5.
In contrast, ${\bar U}(x)$ is a strong increasing function 
in the rigid band approximation \cite{Yamakawa-H}.
Figure \ref{fig:SCgap} (c)
shows the obtained $x$-dependence of the 
$\a_{c,s}$ and $\lambda_E$ for $J/U=0.14$ by using $U=\bar{U}(x)$.
The large $\a_c$ at $x=0$ and that at $x=0.5$
explain the experimental $C_2$ and $C_4$ structure transitions
of LaFeAsO$_{1-x}$H$_x$.
The eigenvalue $\lambda_E$ approximately follows $\a_c$ and shows two peaks 
near the $C_2$ and $C_4$ structure transition points,
due to the strong orbital fluctuations.
Since $T_{\rm c}$ is suppressed by the 
structure transition, the obtained $x$-dependence of $\lambda_E$
would be consistent with the double-dome $T_{\rm c}$.
In contrast, single-dome $T_{\rm c}$ is obtained
in the FLEX approximation in the present model
\cite{suppl}.

Figure \ref{fig:SCgap} (d)-(f) show
the gap functions multiplied by $z_l$
in the band-basis for $x=0$, 0.14 and 0.4, respectively.
At $x\sim0$ and 0.4, 
approximate $s_{++}$-wave states are obtained as shown in (d) and (f), 
due to the strong orbital fluctuations.
At $x\sim0.4$, the gap structure is fully-gapped,
whereas the gap on the e-FS is nodal
at $x\sim0$ due to the competition (cooperation)
of orbital and spin fluctuations
\cite{Saito-loop}.
These $s_{++}$-type gap structures are realized
by taking the $\Delta$-VC into account beyond the Migdal's theorem,
since the attractive interaction due to ${\hat \chi}^c$ in eq. (\ref{eqn:V})
is multiplied by $|{\hat \Lambda}^c(k,q)|^2 \gg1$
\cite{Onari-SCVC,LiFeAs}.
(The $\Delta$-VC can overcome the 
factor $3$ for the spin channel in eq. (\ref{eqn:V})
that favors the $s_\pm$-state.)
The $s_{++}$ state is realized against the strong Coulomb repulsion
due to the retardation effect, since the energy-scale of 
orbital fluctuations is $\sim T$.
The $s_{++}$ state is consistent with the robustness of $T_{\rm c}$
against the randomness in Fe-pnictides
\cite{Sato-imp,Nakajima-imp,Li-imp,Onari-impurity,Yamakawa-impurity,Wang-impurity}.

Figure \ref{fig:SCgap} (e) shows the gap functions for $x=0.14$.
Although the spin fluctuation is stronger
because of the relation $\a_c\ll\a_s$,
the obtained gap structure is very different from the 
so-called $s_\pm$-wave state \cite{Kuroki,Hirschfeld,Chubukov}, 
in which the gaps of the three hole-FSs are the same in sign.
This gap structure is induced by the cooperation of the 
``attractive interaction between h-FS3 and e-FS''
due to orbital fluctuations and 
``repulsive interaction between h-FS1,2 and e-FS''
due to spin fluctuations
\cite{LiFeAs}.
This gap structure may easily change to the $s_{++}$-wave state
by introducing small amount of impurities 
and $e$-ph interaction \cite{Onari-impurity}.

We also performed the similar analysis for SmFeAsO$_{1-x}$H$_x$,
which shows the single-dome $T_{\rm c}$,
by constructing the first-principle tight-binding models.
In Sm-compounds, h-FS3 is very large
due to the shorter anion-height \cite{Hosono-As}, and
the inter- and intra-orbital nesting is improved.
Since the strong orbital fluctuations appear even at $x\sim0.2$,
$\lambda_E$ of SmFeAsO$_{1-x}$H$_x$ becomes larger 
as shown in the inset of Fig. \ref{fig:SCgap} (c),
and the single-dome $T_{\rm c}$ structure is well reproduced.
This result indicates the importance of the $d_{xy}$-orbital FS
to realize higher $T_{\rm c}$.

In summary, we studied the phase diagram of LaFeAsO$_{1-x}$H$_x$
using the SC-VC$_\Sigma$ method,
and predicted that the non-nematic $O_{3z^2-r^2}$ order 
triggers the new $C_4$ isostructural transition at $x\sim0.5$ 
\cite{Hosono-private}.
Also, we obtain the approximate $s_{++}$-wave gap structure 
due to orbital fluctuations for both $x\gtrsim0$ and $x\lesssim0.5$
by taking the $\Delta$-VC into account.
The switch of the dominant quadrupole fluctuations in Fig. \ref{fig:chiQ-U}
gives the minimum structure of $T_{\rm c}$ around $x\sim0.2$.
The non-nematic orbital fluctuations will be
a key ingredient in realizing high-$T_{\rm c}$ ($\sim50$K)
in H-doped La1111, Sm1111, as well as Ca122.

\acknowledgements
We are grateful to 
H. Hosono, J. Yamaura, Y. Murakami, N. Fujiwara, H. Hiraga and S. Iimura
for useful discussions.
This study has been supported by Grants-in-Aid for Scientific 
Research from MEXT of Japan.




\newpage

\section{[Supplemental Material] Numerical Study of LaFeAsO$_{1-x}$H$_x$: 
Comparison Between Different Types of Fluctuation Theories}

\subsection{
A: fluctuation-exchange (FLEX) approximation for LaFeAsO$_{1-x}$H$_x$
}

In the main text, we studied the
models of $Ln$FeAsO$_{1-x}$H$_x$ ($Ln$=La,Sm) based on the 
SC-VC$_\Sigma$ method, in which both the VC and self-energy 
are calculated self-consistently.
Since the nematic and non-nematic orbital orders are induced by the VC, 
experimental $C_2$- and $C_4$-structure transitions are naturally explained.
In addition, orbital-fluctuation-mediated $s_{++}$-wave state is obtained.

The FLEX approximation had been used in the study of 
Fe-based superconductors \cite{Suzuki-H2}.
Considering that the FLEX approximation cannot explain 
the $C_2$ and $C_4$-structure transition due to the neglect of the VC,
it would be incomplete for the study of
the superconductivity realized near the structural QCPs.
With knowledge of this defect, 
we solve the linearized gap equation within the FLEX approximation.

\begin{figure}[!htb]
\includegraphics[width=.6\linewidth]{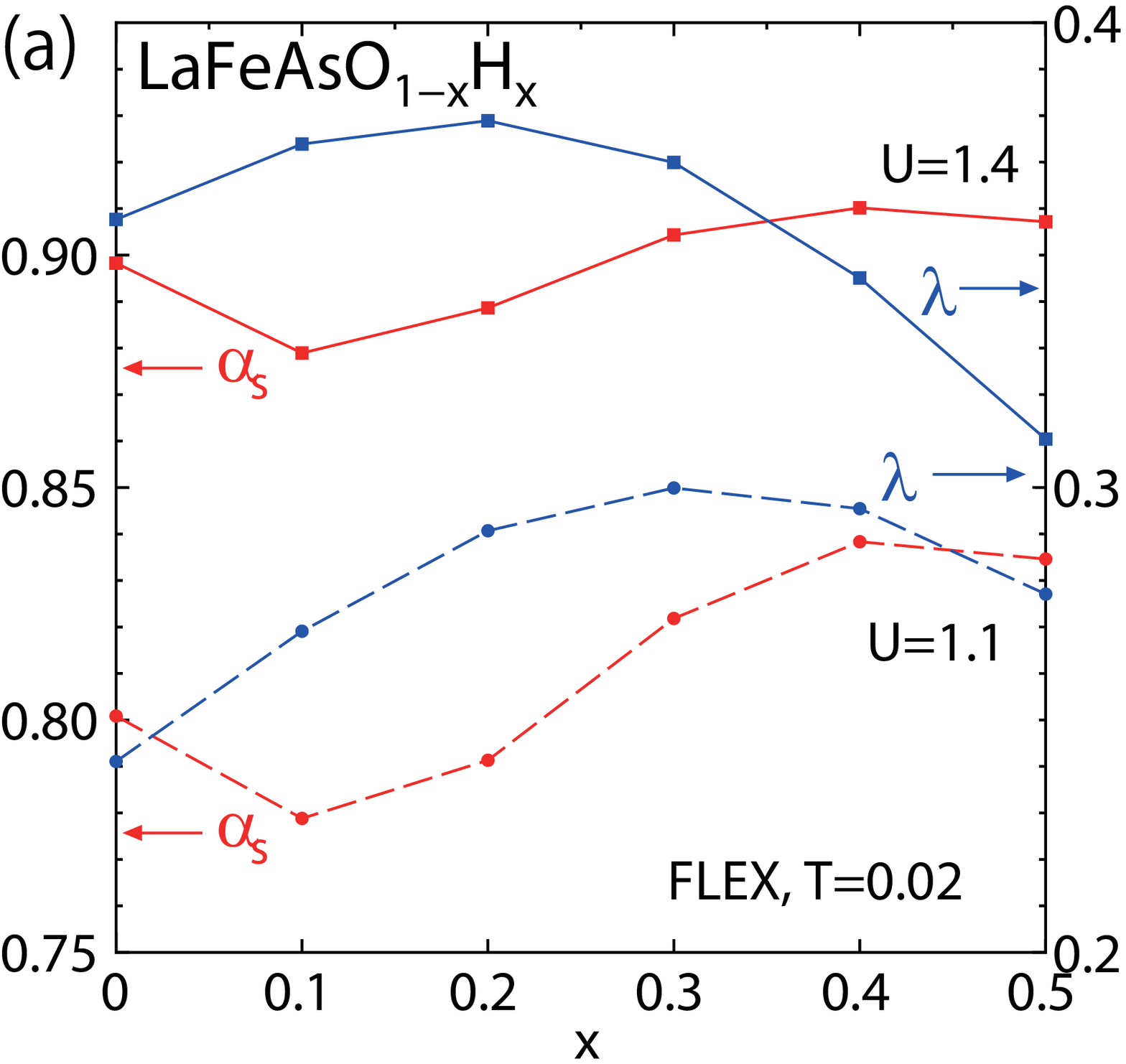}
\includegraphics[width=.6\linewidth]{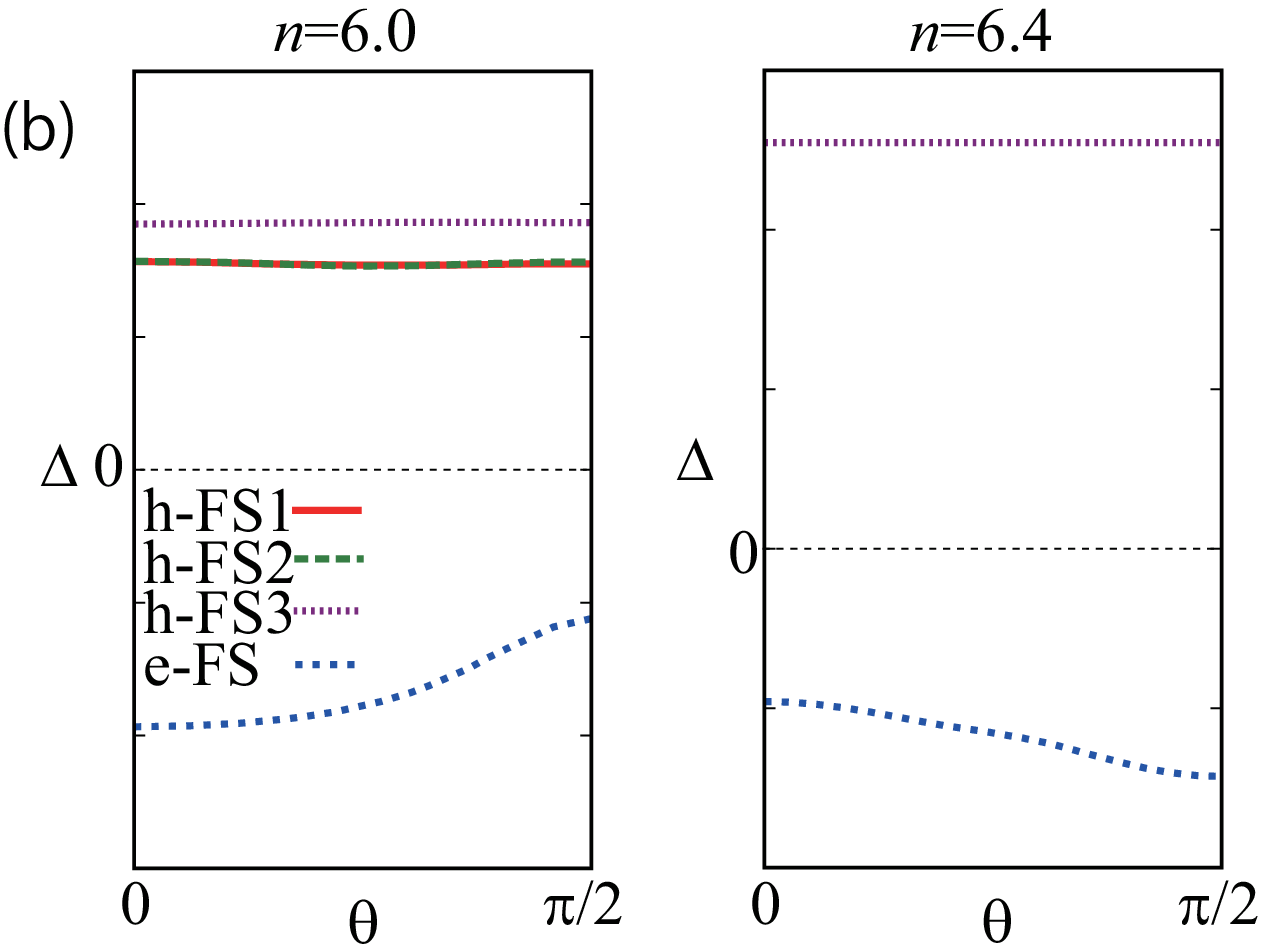}
\caption{(color online)
(a) $x$-dependences of the spin Stoner factor $\a_S$ and 
the eigenvalue $\lambda_E$ given by the FLEX approximation.
(b)(c)  Gap functions given by the FLEX approximation
for $x=0$ and $0.4$ in the case of $U=1.4$.
}
\label{fig:FLEX}
\end{figure}

We use $64\times64$ $\k$-meshes and 512 Matsubara frequencies,
and fix $T=0.02$eV to avoid
artifacts due to the shortage of the $\k$-mesh number.
We fix the ratio $J/U=1/6$.
Figure \ref{fig:FLEX} (a) shows the obtained 
spin Stoner factor $\a_S$ and the eigenvalue of the 
gap equation $\lambda_E$ for $n=6.0\sim6.5$ ($x=0\sim0.5$),
in the case of $U=1.1$ and $1.4$.
Similar result is obtained for $T=0.005$ and $U=1.3$
using $32\times32$ $k$-meshes,
which were used by Suzuki {\it et al}.
\cite{Suzuki-H2}.
The obtained $s_\pm$-wave gap functions are shown in 
Fig. \ref{fig:FLEX} (b) and (c).

The single-dome structure of $\lambda_E$ in Fig. \ref{fig:FLEX} (a)
is different from the numerical result at $\Delta\a=0$ 
in Ref. \cite{Suzuki-H2}, but similar to that of $\Delta\a<-1^\circ$ 
This difference should originate from differences in the models:
Suzuki {\it et al} used the VASP package using a fixed Fe-As length, 
whereas the present authors used the WIEN2k package.

\subsection{
B: self-consistent vertex-correction (SC-VC) method for 
LaFeAsO$_{1-x}$H$_x$
}

In the main text, we analyzed LaFeAsO$_{1-x}$H$_x$ 
using the SC-VC$_\Sigma$ method, in which the self-energy 
correction is incorporated into the SC-VC method.
The obtained strong ferro-quadrupole susceptibilities 
$\chi^Q_{x^2-y^2}(\bm{0})$ and $\chi^Q_{3z^2-r^2}(\bm{0})$ in 
Fig. \ref{fig:chiQ-U} explain the 
$C_2$ and $C_4$ structure transitions at $x\sim0$ and $x\sim0.5$, 
respectively.

\begin{figure}[!htb]
\includegraphics[width=.9\linewidth]{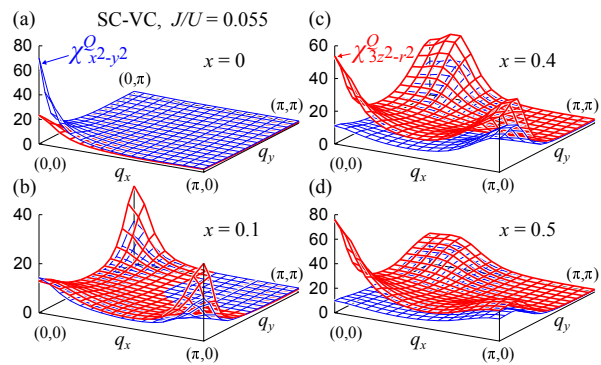}
\caption{(color online)
for (a) $x=0$ ($U=1.39$), (b) $x=0.1$ ($U=1.34$), (c) $x=0.4$ ($U=1.06$)
and (d) $x=0.5$ ($U=0.99$),
obtained by the SC-VC method for $J/U=0.055$.
Stoner factors are $\a_c=0.97$ and $\a_s\sim0.9$.
In the SC-VC method, $\chi_{xz/yz}^Q(q)$ is smaller and not shown.
}
\label{fig:chi-SCVC}
\end{figure}

Here, we show that the essentially similar 
quadrupole fluctuations are obtained by using the SC-VC method,
in which the self-energy is dropped \cite{Onari-SCVC}.
Due to the self-energy, the value of $U$ for a fixed ${\rm max}\{\a_c,\a_s\}$
in the SC-VC$_\Sigma$ method is larger than that in the SC-VC method.
Since the AL-term grows in proportion to $U^4$,
the relation $\a_c>\a_s$ is realized against larger $J/U$ 
in the SC-VC$_\Sigma$ method.

Figure \ref{fig:chi-SCVC} (a)-(d) shows the obtained
$\chi_{3z^2-r^2}^Q(q)$ and  $\chi_{x^2-y^2}^Q(q)$ for $J/U=0.055$:
The results are essentially unchanged for $0<J/U<0.06$.
For each $x$, we choose $U$ so that
the charge Stoner factor satisfy $\a_c=0.97$.
At $x=0$ in (a) and $x=0.1$ in (b), 
we obtain large peak of $\chi_{x^2-y^2}^Q({\bm 0})$,
and its divergence corresponds to the $C_2$ structure transition.
At $x=0.4$ in (c) and $x=0.5$ in (d),
we obtain the divergent peak of $\chi_{3z^2-r^2}^Q({\bm 0})$,
which corresponds to the $C_4$ isostructural transition.
Thus, both $C_2$ and $C_4$ structure transitions
in LaFeAsO$_{1-x}$H$_x$ at $x\sim0$ and $x\sim0.5$ respectively
are explained by the SC-VC method,
meaning that the self-energy correction is not essential for them.

In addition, strong antiferro-quadrupole susceptibilities
$\chi_{3z^2-y^2}^Q({\bm Q})$ and $\chi_{x^2-y^2}^Q({\bm Q})$
appears for $x\ge0.1$ in Fig. \ref{fig:chi-SCVC} (b)-(d).
On the other hand, $\chi_{xz/yz}^Q({\bm Q})$ remains small, 
although it is strongly enhanced in the SC-VC$_\Sigma$ method 
shown in Fig. \ref{fig:chiQ-U} (a).
It is considered that this discrepancy originates from the
neglect of the self-energy in the SC-VC method:
In the SC-VC$_\Sigma$ method, the strong $d_{xy}$-orbital spin susceptibility 
$\chi^s_{4,4;4,4}$ in the RPA are suppressed by the $d_{xy}$-orbital self-energy.
Due to this negative feedback effect,
$\chi^s_{4,4;4,4}$ is comparable to $\chi^s_{2,2;2,2}$ and $\chi^s_{3,3;3,3}$
in the SC-VC$_\Sigma$ method, and then $\Phi_{2(3)}^c\sim \Phi_4^c$.
For this reason, $\chi_{xz/yz}^Q(\Q)$ is enlarged by large $\Phi_{2(3)}^c$.


In the previous SC-VC study for $x=0.1$ \cite{Onari-SCVC},
we have interested in the developments of 
$\chi_{x^2-y^2}^Q(q)$ and $\chi_{xz,yz}^Q(q)$,
so we have dropped $X_{l,l,;4,4}^c(q)$ and $X_{4,4;l,l}^c(q)$.
In the present study, we include $X_{l,l;4,4}^c(q)$, 
and find that $\chi_{3z^2-r^2}^Q(q)$ is also strongly enhanced 
due to large $X_{4,4;4,4}^c(q)$.
However, large $\chi_{3z^2-r^2}^Q$ for $x=0\sim0.1$ 
in Fig. \ref{fig:chi-SCVC} (b) is found to be over-estimated
due to the absence of the self-energy, 
as confirmed by the numerical result of the SC-VC$_\Sigma$ method
in Fig. \ref{fig:chiQ-U} (a).
Except for that,
the obtained results given by the SC-VC method
are similar to those by the SC-VC$_\Sigma$ method,
and therefore they are reliable.

\subsection{
C: SC-VC$_\Sigma$ method for $J/U=0.12$
}

\begin{figure}[!htb]
\includegraphics[width=.95\linewidth]{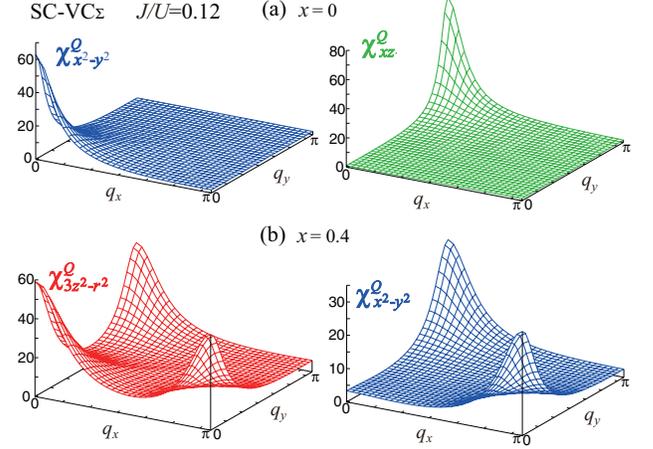}
\caption{(color online)
$\chi_\gamma^Q(\q)$ 
given by the SC-VC$_\Sigma$ method for $J/U=0.12$:
(a) $\gamma=x^2-y^2$ and $\gamma=xz$ for $x=0$ ($U=2.05$), and
(b) $\gamma=3z^2-r^2$ and $\gamma=x^2-y^2$ for $x=0.4$ ($U=1.62$).
In both cases, $\a_c=0.97$.
}
\label{fig:chiQ012}
\end{figure}

In Fig. \ref{fig:chiQ-U} (a) and (b) of the main text,
we show the quadrupole susceptibilities 
given by the SC-VC$_\Sigma$ method in the case of $J/U=0.14$.
Here, we perform the same calculation for $J/U=0.12$,
and show the obtained results in Fig. \ref{fig:chiQ012}.
As for the ferro-quadrupole fluctuations,
strong development of $\chi_{x^2-y^2}^Q(\0)$ at $x=0$ and 
that of $\chi_{3z^2-y^2}^Q(\0)$ at $x=0.4$ are obtained in 
Fig. \ref{fig:chiQ012}, 
consistently with the results of $J/U=0.14$.
Also, large antiferro-quadruple susceptibility 
$\chi_{xz,yz}^Q(\Q)$ at $x=0$ is obtained in both $J/U=0.12$ and $0.14$,
whereas $\chi_{3z^2-r^2}^Q(\Q)$ at $x=0.4$ is relatively small for $J/U=0.14$.
As shown in Fig. \ref{fig:SCgap} (c),
the $x$-dependence of the eigenvalue $\lambda_E$ for $J/U=0.12$
is similar to that for $J/U=0.14$.
Thus, qualitative results of the SC-VC$_\Sigma$ method 
are unchanged for $J/U=0.12\sim0.14$.

\subsection
{D: Expressions of 
$\chi^Q_{x^2-y^2}(\bm{0})$ and $\chi^Q_{3z^2-r^2}(\bm{0})$ for $J>0$
}

In the main text, 
we present the analytic expression of 
$\chi^Q_{x^2-y^2}(\bm{0})$ and $\chi^Q_{3z^2-r^2}(\bm{0})$
in Eqs. (\ref{eqn:ap-chiQx2y2}) and (\ref{eqn:ap-chiQz2}), respectively,
in the case of $J=0$ for simplicity.
Here, we present their expressions for finite $J$:
\begin{eqnarray}
&&\chi_{x^2-y^2}^Q({\bm0}) = 2\Phi_2^c(1-(U-5J)\Phi_2^c)^{-1},
\label{eqn:ap-chiQx2y2-2} \\
&&\chi_{3z^2-r^2}^Q({\bm0}) = b'(1-a'U\Phi_4^c)^{-1},
\label{eqn:ap-chiQz2-2}
\end{eqnarray}
where $\displaystyle a'= \frac1U
\frac{5(U-5J)(U-2J)\Phi_2^c -U}{(3U-5J)\Phi_2^c +1}$
and $\displaystyle b'=\frac12
\frac{2\Phi_4^c+\Phi_2^c+15(U-2J)\Phi_2^c\Phi_4^c}{(3U-5J)\Phi_2^c +1}$.

Equation (\ref{eqn:ap-chiQx2y2-2})
had been already given in Ref. \cite{Ohno-SCVC}.
Thus, $\chi_{x^2-y^2}^Q({\bm0})$ diverges when $\Phi_2^c =(U-5J)^{-1}$,
and $U-5J$ is positive when the relation $J/U=0.12\sim0.15$
predicted by the first principle study by Miyake {\it et al}
(J. Phys. Soc. Jpn. {\bf 79}, 044705 (2010)) is correct.
According to the expression of $a'$,
we find that $a'$ is positive for $\Phi_2^c > [5(1-5J/U)(U-2J)]^{-1}$
in the case of $J/U<1/5$.
Then, $\chi_{3z^2-r^2}^Q({\bm0})$ diverges when $\Phi_4^c =(a'U)^{-1}$.
Therefore, the discussions in the main text below 
Eqs. (\ref{eqn:ap-chiQx2y2}) and (\ref{eqn:ap-chiQz2})
is valid even for finite $J$.


\begin{thebibliography}{99}
\bibitem{comment}
For $x>0.5$, in addition to the $c$-axis length change, 
Fe-layers and As-layers slide alternatively at $T_S$ \cite{Hosono-private}.
The latter may be due to band-JT effect since the 
lattice deformation is very large.
\bibitem{comment2}
The nonzero matrix elements of ${\hat O}_\gamma$ 
with respect to $2\sim4$ orbitals 
are $O_{x^2-y^2}^{2,2}=- O_{x^2-y^2}^{3,3}=1$,
$2 O_{3z^2-r^2}^{2,2}=2 O_{3z^2-r^2}^{3,3}=- O_{3z^2-r^2}^{4,4}=1$,
and $ O_{xz}^{3,4}= O_{xz}^{4,3}=1$
\cite{Kontani-RPA}.
\end{thebibliography}
\end{document}